\documentclass[preprint,aps]{revtex4}
\usepackage{graphicx}%
\usepackage{dcolumn}
\usepackage{amsmath}
\usepackage{longtable}
\begin{document}
\title
{ A Mixed Basis Density Functional Approach for Low Dimensional Systems
  with B-splines }
\author
{Chung-Yuan Ren$^{a,\dagger}$, Chen-Shiung Hsue$^{b}$, and
 Yia-Chung Chang$^{c,d}$  }
\affiliation
{ $^{a}$ Department of Physics, National Kaohsiung Normal University,
Kaohsiung 824, Taiwan  	\\
$^{b}$ Department of Physics, National Tsing Hua University,
Hsinchu 300, Taiwan \\
$^{c}$ Research Center for Applied Sciences, Academia Sinica, Taipei 115, Taiwan \\
$^{d}$ {Department of Physics, National Cheng-Kung University, Tainan, Taiwan 701}\\
$\dagger$ {\it E-mail address:} cyren@nknu.edu.tw }

\begin{abstract}
A mixed basis approach based on density functional theory is employed  
for low dimensional systems. The basis functions are taken to be plane waves for the periodic
direction multiplied by B-spline polynomials in the non-periodic direction.
B-splines have the following advantages: (1) the associated matrix elements
are sparse, (2) B-splines possess a superior treatment of derivatives,
(3) B-splines are {\it not} associated with atomic positions when
the geometry structure is optimized, making the geometry optimization easy to implement.
With this mixed basis set we can directly
calculate the total energy of the system instead of using the
conventional supercell model with a slab sandwiched between vacuum regions.
A generalized Lanczos-Krylov iterative method is implemented for the
diagonalization of the Hamiltonian matrix.
To demonstrate the present approach, we apply it to study the
C(001)-(2$\times$1) surface with the norm-conserving pseudopotential,
the n-type $\delta$-doped graphene, and graphene nanoribbon with
Vanderbilt's ultra-soft pseudopotentials.
All the resulting electronic structures were found to be
in good agreement with those obtained by the VSAP code,
but with a reduced number of basis. 
\end{abstract}
\maketitle
\section{INTRODUCTION}
The electronic properties of low-dimensional systems are fundamentally
different from those in higher dimensions due to their unusual collective
excitations. Nowadays, 1-dimensional (1D) or 2-dimensional (2D) materials
can be easily fabricated due to the emerging nanotechnology, which leads to
intensive exploration of low-dimensional systems for material innovation.

\begin{figure}[h]
        \caption{ } \label{fig1}
\includegraphics{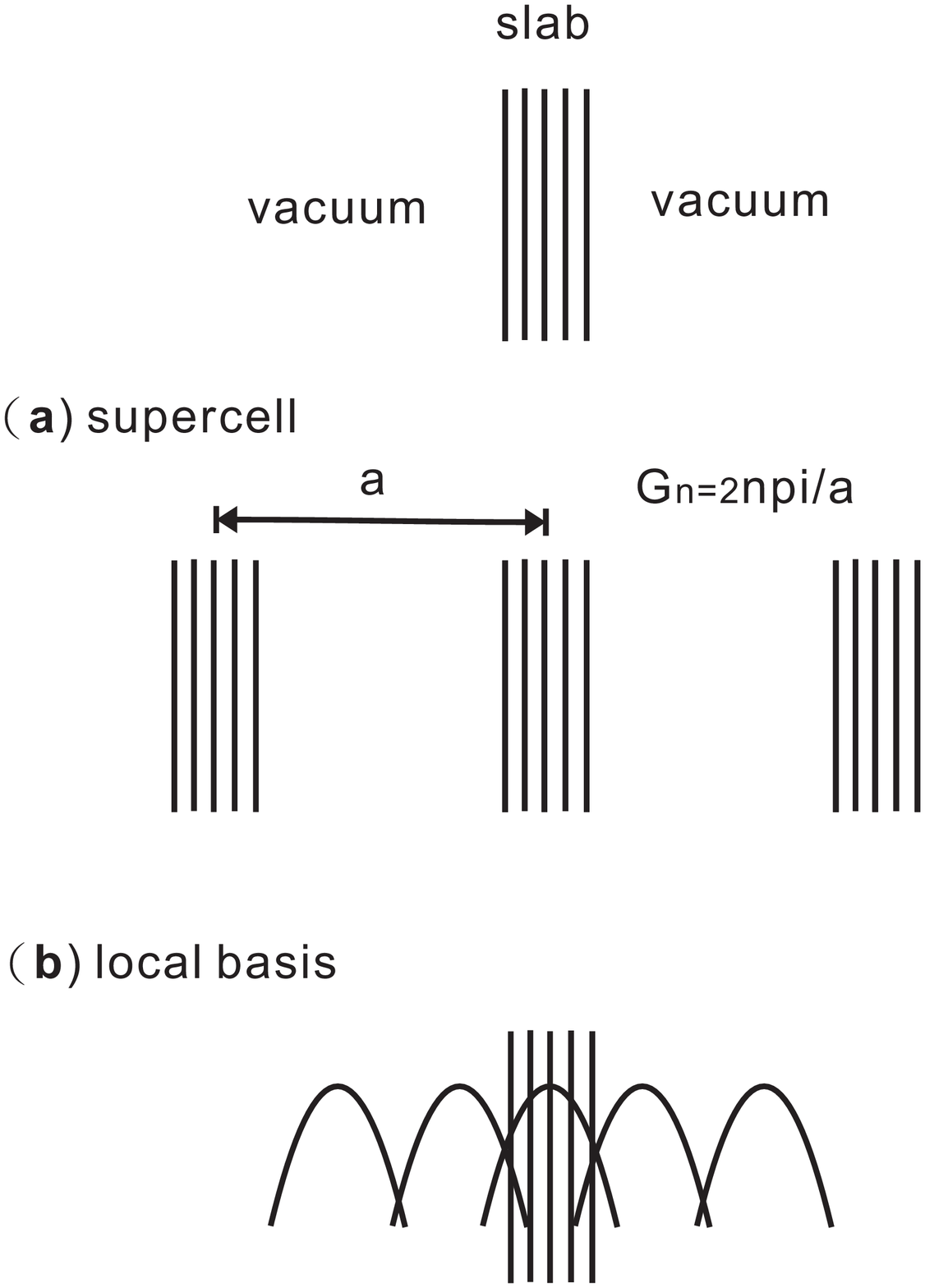}
\end{figure}

First-principles methods based on the density functional theory have been
used extensively in investigating the electronic structures and properties of
solids. In most calculations, three-dimensional (3D) plane waves are used as
basis functions, which are very suitable for systems that are
periodic in three dimensions. Because of its easy implementation and the fact
that  the convergence of the calculation can be checked
systematically,  plane waves are often employed to expand
the wavefunction even along the non-periodic direction in low-dimensional systems via the use of 
supercell. In this approach, the physical low-dimensional system is treated as a fully
3D periodic system by introducing some artificial vacuum space to separate
the periodic replica along the direction, in which the system should be
consider as nonperiodic (Fig. \ref{fig1}(a)).
The drawback of this approach is the requirement of large thickness of the vacuum layer such that the interactions between the adjacent slabs are negligible, and therefore increases the number
of the plane wave along that direction. More seriously, for charged systems (e.g. charged defects), the rather long-range tail of the Coulomb potential inevitably requires an extremely large separation of the two slabs and makes the calculation impractical.

Li and Chang have previously introduced a mixed planar basis which is the product of
2D plane waves in the periodic directions and 1D Gaussian function along the
growth direction $z$ for the first-principles calculations \cite{LC}-\cite{CL}.
The use of this mixed basis resumes the layer-like local geometry which appears
in surfaces. This mixed basis is suitable for
describing electronic properties of low-dimensional systems including surfaces,
interface, and superlattices. Moreover, because one can calculate the total
energy for an isolated slab instead of using a supercell consisting of alternating slab
and vacuum regions, the physical quantity of a surface, such as the work function can
be immediately obtained without any correction.

However, the Gaussian-type orbital basis falls off too rapidly for large $z$,
i.e., it behaves like $e^{-z^2}$ rather than $e^{-z}$ for the hydrogenic
orbitals. Increasing the number of the orbitals makes no guarantee of numerical
stability and good convergence.  Furthermore, the Gaussian orbital is an
atom-centered basis and will move accordingly in geometry
optimization process, making the coding more difficult and harder to maintain.
Inspired by previous works \cite{JBS}-\cite{RJH}, in the present work
we replace the Gaussian orbitals by B-splines \cite{deBoor}.
B-splines are highly localized and piecewise polynomials within
prescribed break points which consist of a sequence numbers called knot
sequence. B-splines have several advantages over Gaussian basis functions:
(1) the relevant matrices are sparse. (2) B-splines are
superior to traditional methods in the treatment of derivatives.
(3) B-splines possess good flexibility to represent a rapidly varying
wavefunction accurately with the knots being arbitrarily chosen to
have an optimized basis. (4) Finally, B-splines are independent of
atomic positions, so the geometry optimization can be easily implemented.

We also implement a generalized Lanczos-Krylov iterative method for the
diagonalization of the Kohn-Sham Hamiltonian.
It offers a stable and promising way to find the low-lying
eigenvalues of a real matrix with a very large dimension via an
iterative scheme.

To test the application of the present approach, we will choose the
C(001)-(2$\times$1) surface, n-type $\delta$-doped graphene, and
the graphene nanoribbon as three examples. We perform the band structure
calculation for the former case using
the norm-conserving pseudopotential (NCPP) \cite{GTH}.
For the latter two cases, we adopt
Vanderbilt's ultra-soft pseudopotentials (USPP) \cite{DV}
with modest energy cutoff.
It is found that the band structures are all in good agreement with those by the popular
VASP code \cite{KJ,KF}, but the number of basis functions needed is reduced.
\section{METHOD OF CALCULATION}
\subsection{B-splines}
B-splines form a convenient basis set with good flexibility. Here we
briefly summarize the B-spline formalism.
More details can be found in Ref. \cite{deBoor}.
In general, B-spline of order $\kappa$ consists
of positive polynomials of degree $\kappa-1$, over $\kappa$
adjacent intervals. These
polynomials vanish everywhere outside the subintervals $\tau_i < z <
\tau_{i+\kappa} $.
The B-spline basis set given by the order $\kappa$ and the knot sequence $\{ \tau_i\}$
is generated by the following relation :
\begin{equation}
B_{i,\kappa}(z)=\frac{z-\tau_i}{\tau_{i+\kappa-1}-\tau_i}B_{i,\kappa-1}(z)+\frac{\tau_{i+\kappa}-x}{\tau_{i+\kappa}
-\tau_{i+1}}B_{i+1,\kappa-1}(z),
\end{equation}
with
\begin{equation}
B_{i,1}(z)= \left \{ \begin{array}{ll}
                1,      & \tau_i \leq x < \tau_{i+1}  \\
                0,      & {\rm otherwise \ .}
                \end{array}
                \right.
\end{equation}
The first derivative of the B-spline of order $\kappa$ is given by
\begin{equation}
\frac{d}{dz}B_{i,\kappa}(z)=\frac{\kappa-1}{\tau_{i+\kappa-1}-\tau_i}B_{i,\kappa-1}(z)-\frac{\kappa-1}
{\tau_{i+\kappa}-\tau_{i+1}}B_{i+1,\kappa-1}(z).
\end{equation}
Since all the lower order B-splines are the simple polynomials,
it is obvious that
B-splines are superior to traditional methods in the treatment
of derivatives.
Figure 2 shows the B-splines generated from an equal
distance and exponential type knot sequence.
Clearly, B-splines possess good flexibility to accurately represent any localized 
function of $z$ by suitably increasing the density of the knot
sequence where it varies rapidly.

\subsection{Relevant matrix elements within B-spline basis}
\subsubsection{NCPP scheme}
In the NCPP scheme, the minimization of total energy subject to the constraint
that the wave functions $\phi_i$ are orthogonal yields
\begin{equation}
H|\phi_i>=\epsilon_i|\phi_i> \label{KHeq}
\end{equation}
where
\begin{equation}
H=- \nabla^2\ + V_{pp} + V_H+ V_{xc}\ .
\end{equation}
Here, $V_{pp}$, $V_H$, and $V_{xc}$ denote the pseudopotential, Hartree potential,
and exchange-correlation potential, respectively.

The present mixed basis used to expand $\phi_i$
is defined as
\begin{equation}
< {\bf r } | { \bf \ k_\parallel +  G_\parallel } ; j, \kappa > \
=
\frac{1}{\sqrt{A}}\
e^{i( {\bf k_\parallel +  G_\parallel} ) \cdot { \bf r_{\parallel}} }
\ B_{j,\kappa}(z)
\end{equation}
where
$ {\bf G_\parallel} $
denotes an in-plane reciprocal lattice vector,  and
$ {\bf r_{\parallel}= (x, y)} $ is the projection of $ \bf r$ in the x-y plane.
$ {\bf k_\parallel} $ is the in-plane wave vector.
$A$ is the surface area of the system.

The overlap matrix elements between two basis states are given by
\begin{equation}
< {\bf k_\parallel + G_\parallel} ; i, \kappa \
|\ {\bf k_\parallel +  G'_\parallel } ; i', \kappa
> \
= \
< \ B_{i,\kappa}
| \ B_{i',\kappa}>\ \delta_{ {\bf G_\parallel, G'_\parallel } }
\end{equation}
where
\begin{equation}
< \ B_{i,\kappa}
| \ B_{i',\kappa}>\ = \int dz \ B_{i,\kappa}(z)\ B_{i',\kappa}(z)
\end{equation}
is an integration of local polynomials with bounded support and vanishes
unless the condition $|i-i'|\le \kappa$
is fulfilled.

The kinetic energy matrix elements are given by
\begin{align}
\lefteqn{
< {\bf k_\parallel +  G'_\parallel } ; i, \kappa
\ |\  - \nabla^2\  |\
 { \bf k_\parallel +  G_\parallel} ; i', \kappa
>
}
\nonumber\\
=& [ \
< \ B_{i,\kappa} | -\frac{\partial^2}{\partial z^2}| \ B_{i',\kappa}>\
+
< \ B_{i,\kappa} |\ B_{i',\kappa}>\
({\bf k_\parallel + G_\parallel})^2 \
]\
\delta_{ {\bf G_\parallel, G_\parallel} }
\nonumber\\
=& [ \
< \ {B'}_{i,\kappa} | \ {B'}_{i',\kappa}>\
+
< \ B_{i,\kappa} |\ B_{i',\kappa}>\
( {\bf k_\parallel + G_\parallel} )^2 \
]\
\delta_{ {\bf G_\parallel, G'_\parallel} } \ .
\end{align}

$ {B'}_{i,\kappa}(z)$, the derivatives of $ B_{i,\kappa}(z)$,
can be expressed as a linear combination of
$ B_{i,\kappa-1}(z)$ with order $\kappa -1$ and is continuous
across the knot sequence.

The local part of $V_{pp}$ on each atomic site with species $\sigma$
concerned here can be written as
\begin{equation}
V_{\text{loc}}^{\sigma} ({\bf r}) =
\ -\  \frac{Z^{\sigma}}{r}\ \text{erf}\left( \frac{r}{R_c^{\sigma}}\right)
+ \sum_i A_i^{\sigma} e^{- a_i^{\sigma}r^2}.
\end{equation}

The first term on the right hand side of the above equation
will be referred to as the core term due to
the core charge distribution
\[
\rho_c( r) \ = \frac{Z^{\sigma}}{\pi^{\frac{3}{2}} {R_c^{\sigma}}^3}\
e^{- \frac{r^2}{{R_c^{\sigma}}^2}} \ .
\]

The local pseudopotential of the crystal is then given by
\begin{equation}
V_{\text{LOC}} ({\bf r}) = \sum_{{\sigma,\bf R^\sigma}}
V_{\text{loc}}^{\sigma} ({\bf r}- {\bf R^\sigma}) \ ,
\end{equation}
where ${\bf R^\sigma}$ denotes the
position of each atom with species $\sigma$.
The matrix elements for the local pseudopotential of the crystal
excluding the core term,
$
V'_{\text{LOC}}
$,
are given by
\begin{multline}
 <\ {\bf k_\parallel} + {\bf G_\parallel} ; i, \kappa \
 |\  V'_{\text{LOC}}\ |
\ {\bf k_\parallel} + {\bf G'_\parallel} ; i', \kappa > \
=
\\
 \sum_{\sigma, R_z\in SUC}
I(i, \kappa; a_i^{\sigma}, R_z^\sigma; i' , \kappa)
\ e^{-i\boldsymbol{\Delta}{\bf G_\parallel}
\cdot {\bf R_\parallel^\sigma}}
\ {\tilde V}_{\text{loc}}^{\sigma} (\boldsymbol{\Delta}{\bf G_\parallel})
\end{multline}
where
$
\boldsymbol{\Delta}{\bf G_\parallel}
={\bf G_\parallel}- {\bf G'_\parallel}
$
and
\begin{align}
{\tilde V}_{\text{loc}}^{\sigma} (\boldsymbol{\Delta}{\bf G_\parallel)}
=&
\iint d^2 {\bf r_{\parallel}}\  e^{-i(
\boldsymbol{\Delta}{\bf G_\parallel}\cdot {\bf r_{\parallel}}- a_i^{\sigma}
 r_{\parallel}^2)}
\nonumber\\
=& \sum_i\left(\frac{A_i^{\sigma}}{A_c}\right) \frac{\pi}{a_i^{\sigma}}
e^{-\ \frac{ \boldsymbol{\Delta}{\bf G_\parallel^2}}{4 a_i^{\sigma}}}
\ .
\end{align}
$ A_c$ is the area of the surface unit cell (SUC).
$I$ is given by
\[
I(i, \kappa; a_i^{\sigma} , R_z^{\sigma}; i', \kappa) = \int dz
\ B_{i,\kappa}(z)B_{i',\kappa}(z) e^{- a_i^{\sigma}(z-R_z^{\sigma})^2} \ .
\]

The atomic nonlocal pseudopotential associated with species $\sigma$
used in the present work is in the Kleinman-Bylander form \cite{KB}, viz.,
\begin{equation}
V_{\text{nl}}^{\sigma}({\bf r}) = \sum_{nlm} E_{nl}^{\sigma}
|\beta_{n,lm}^{\sigma}>
<\beta_{n,lm}^{\sigma}|\ . \label{KBeq}
\end{equation}
The projector $\beta_{n,lm}^{\sigma}$
is a radial function multiplied by an angular momentum eigenfunction
$Y_{lm}(\boldsymbol{\Omega})$.
For the norm-conserving pseudopotential by Goedecker, Teter, and Hutter
(GTH) \cite{GTH}, there are two $s$-channels and one $p$-channel for the second
row element,
\begin{equation}
\beta_{1,00}^{\sigma}({\bf r}) = C_{1,0}^{\sigma}
\ e^{- \alpha_{0}^{\sigma}r^2}\ r^0 Y_{00}(\boldsymbol{\Omega}) \nonumber 
\end{equation}
\begin{equation}
\beta_{2,00}^{\sigma}({\bf r}) = C_{2,0}^{\sigma}
\ e^{- \alpha_{0}^{\sigma}r^2}\ r^2 Y_{00}(\boldsymbol{\Omega})
\end{equation}
\begin{equation}
\beta_{1,1m}^{\sigma}({\bf r}) = C_{1,1}^{\sigma}
\ e^{- \alpha_{1}^{\sigma} r^2}\ r^1 Y_{1m}(\boldsymbol{\Omega})\ \nonumber.
\end{equation}
Since the $d$-like pseudopotential is chosen as the local pseudopotential,
only $l=0$ and $l=1$ are considered here.
The nonlocal pseudopotential of the crystal is
\begin{equation}
V_{\text{NL}} ({\bf r}) = \sum_{\sigma, {\bf R^\sigma}}
V_{\text{nl}}^{\sigma}  ({\bf r} - {\bf R^\sigma}) \label{nlcry}
\end{equation}
and the matrix elements of
$ V_{\text{NL}}$
are given by
\begin{eqnarray}
<\ {\bf k_\parallel} + {\bf G_\parallel} ; i, \kappa
&|& V_{\text{NL}}\ |
{\bf k_\parallel} + {\bf G'_\parallel} ; i', \kappa  >
\ \nonumber
\\
=
&&
\sum_{\sigma,{\bf R^\sigma}}\sum_{nlm}
<\ {\bf k_\parallel} + {\bf G_\parallel} ; i, \kappa
|\beta_{n,lm}^{{\sigma},\bf R^{\sigma}} >
E_{nl}^{\sigma}
<\beta_{n,lm}^{\sigma,{\bf R^{\sigma}}} |\
\ {\bf k_\parallel} + {\bf G'_\parallel} ; i', \kappa  >
\label{inpro}
\end{eqnarray}
where
$
\beta_{n,lm}^{\sigma,{\bf R^{\sigma}}}\equiv
\beta_{n,lm}^{\sigma}({\bf r} - {\bf R^{\sigma}})
$.
By the representation of $r^lY_{lm}(\boldsymbol{\Omega})$
in Cartesian coordinates for $l = 0$ and $l = 1$,
\begin{align*}
r^0Y_{00}(\boldsymbol{\Omega})=&\  \frac{1}{\sqrt{4 \pi}}
, \quad \
&r^1Y_{1-1}(\boldsymbol{\Omega})= &\quad \sqrt{\frac{3}{8 \pi}}\ (x-iy)
\\
r^1Y_{10}(\boldsymbol{\Omega})=&\  \sqrt{\frac{3}{4 \pi}}z
, \quad
&r^1Y_{11}(\boldsymbol{\Omega})=&\ - \sqrt{\frac{3}{8 \pi}}\ (x+iy)
\\
r^2Y_{00}(\boldsymbol{\Omega})=&\  \frac{1}{\sqrt{4 \pi}}(x^2+y^2+z^2)
\end{align*}
we obtain
\begin{multline}
<\ {\bf k_\parallel} + {\bf G_\parallel} ; i, \kappa
|  V_{\text{NL}} \ |
\ {\bf k_\parallel} + {\bf G'_\parallel} ; i', \kappa  >
\ =   \\
\ \frac{1}{A_c}
\sum_{\sigma,{\bf R^\sigma} \in SUC}\sum_{nlm}
e^{-i\boldsymbol{\Delta}{\bf G_\parallel}\cdot
{\bf R_\parallel^\sigma}}
\nonumber
E_{nl}^{\sigma}
\left[ C_{n,l}^{\sigma} S_{n,lm}^{\sigma}
(i, \kappa;\ {\bf k_\parallel} + {\bf G_\parallel}; R_z^{\sigma}, 
\alpha_l^{\sigma})
\right]^*
\\
\times \left[ C_{n,l}^{\sigma} S_{n,lm}^{\sigma}
(i', \kappa;\ {\bf k_\parallel} + {\bf G'_\parallel}; R_z^{\sigma}, 
\alpha_l^{\sigma})
\right]
\end{multline}
where
\begin{eqnarray*}
\ && S_{n,lm}^{\sigma}
(i, \kappa;\ {\bf k_\parallel} + {\bf G_\parallel}; R_z, \alpha_l^{\sigma})  
=
\left \{ \begin{array}{ll}
	Z_{1,lm}Q_{1,lm}, & \mbox{$n=1$} \\
        Z_{1,00}Q_{2,00}+ Z_{2,00}Q_{1,00}, & \mbox{$n=2$}
	 \end{array}
\right.
\end{eqnarray*}
with the factors
$ Z_{n,1m}$ and $ Q_{n,lm}$ given explicitly as
\begin{align*}
Z_{1,00} =&\
\int dz
\ B_{i,\kappa}(z)\
\ e^{- \alpha_0^{\sigma}(z-R_z^\sigma)^2}
\\
Z_{1,1\pm1} =&\
\int dz
\ B_{i,\kappa}(z)\
\ e^{- \alpha_1^{\sigma}(z-R_z^\sigma)^2}
\\
Z_{1,10} =&\
\int dz
\ B_{i,\kappa}(z)\
(z-R_z^\sigma)\
\ e^{- \alpha_1^{\sigma}(z-R_z^\sigma)^2}
\\
Z_{2,00} =&\
\int dz
\ B_{i,\kappa}(z)\
(z-R_z^\sigma)^2 \
\ e^{- \alpha_0^{\sigma}(z-R_z^\sigma)^2}
\\
 Q_{1,l0}
= &
\iint d^2 {\bf r_\parallel}
e^{i({\bf k_\parallel} + {\bf G_\parallel} ) \cdot {\bf r_{\parallel}} }
\ e^{- \alpha_l^{\sigma}
{\bf r_\parallel}^2}
\\ =&
\ \frac{\pi}{\alpha_l^{\sigma}} \ e^{- \frac{|{\bf k_\parallel} + 
{\bf G_\parallel}|^2 } {4 \alpha_l^{\sigma}}}
\\
 Q_{1,1\pm1}
= & \ -
\iint d^2 {\bf r_\parallel}
\left(
\frac{x\pm iy}{\sqrt{2}}
\right)\
e^{i({\bf k_\parallel} + {\bf G_\parallel} ) \cdot {\bf r_{\parallel}} }
\ e^{- \alpha_1^{\sigma}
{\bf r_\parallel}^2}
\\ =& -
\left(
\frac{1}{\sqrt{2}}  \frac{k_x + G_x}{2 \alpha_1^{\sigma}} \pm
\frac{i}{\sqrt{2}}  \frac{k_y + G_y}{2 \alpha_1^{\sigma}}
\right)\
\frac{\pi}{\alpha_1^{\sigma}}
\ e^{- \frac{|{\bf k_\parallel} + {\bf G_\parallel}|^2 }
{4 \alpha_1^{\sigma}}}
\\
Q_{2,00}
= & \
\iint d^2 {\bf r_\parallel}
(x^2+y^2)
e^{i({\bf k_\parallel} + {\bf G_\parallel} ) \cdot {\bf r_{\parallel}} }
\ e^{- \alpha_0^{\sigma}
{\bf r_\parallel}^2}
\\ =&
\left(
\frac{1}{\alpha_0^{\sigma}}- \frac{|{\bf k_\parallel} + {\bf G_\parallel}|^2}
 {4 {\alpha_0^{\sigma}}^2}
\right)\
\frac{\pi}{\alpha_0^{\sigma}}
\ e^{- \frac{|{\bf k_\parallel} + {\bf G_\parallel}|^2 }
{4 \alpha_0^{\sigma}}} \ .
\end{align*}

The total charge distribution $\rho$
is defined as the sum of the core charge distributions
for all atoms in the sample $\rho_c$, plus the electronic charge
distributions $\rho_e$,
\begin{equation}
\rho({\bf r})= \rho_c({\bf r})+ \rho_e({\bf r})
= \rho_c({\bf r})+ \sum_i |\phi_i({\bf r})|^2 \ .
\end{equation}
The remainder of the crystal potential includes the Hartree potential
$V_H$ due to the electron charge distribution,
the exchange-correlation potential
$V_{xc}$, and the core term of the local pseudopotential
$V^{(C)}_{\text{LOC}}$ omitted in the above.
Let
$V^{scf}\equiv V_{H}+ V_{xc}+ V^{(C)}_{\text{LOC}}\ .$
This potential is local and periodic in the plane.
The exchange-correlation potential used here are deduced from the Monte Carlo
results calculated by
Ceperley and Alder\cite{CA} and parametrized by Perdew and Zunger\cite{PZ}.
We write
\begin{equation}
V^{scf}({\bf r})
=\sum_{{\bf g_\parallel}}
V^{scf}(z,{\bf g_\parallel})
\ e^{i {\bf g_\parallel}\cdot{\bf r_{\parallel}}} \ .
\end{equation}
The matrix elements of
$V^{scf}$
are given by
\begin{equation}
<{\bf k_\parallel} + {\bf G_\parallel} ; i, \kappa
| V^{scf} |
{\bf k_\parallel} + {\bf G'_\parallel} ; i', \kappa
> \
= \int dz
\ B_{i;\kappa}(z)\ B_{i';\kappa}(z)
V^{scf}(z,\boldsymbol{\Delta}{\bf G_\parallel}) \ .
\end{equation}
With the use of the 2D Fourier transformation of $1/r$
\[
\frac{1}{r}
= 4 \pi \iint \frac{d^2 {\bf k_\parallel}}{ (2 \pi)^2}
\ e^{i{\bf k_\parallel}\cdot {\bf r_{\parallel}}}\
\frac{1}{2 k_\parallel}\  e^{- k_\parallel |z|}
\]
the Coulomb potential due to the total charge distribution is given by
\begin{align}
V^{(C)}=& V^{(C)}_{\text{LOC}}+ V_{H}
= \ \iiint d^3 {\bf r'} \frac{\rho({\bf r'})}{|{\bf r}- {\bf r'}|}
\nonumber\\
= & \sum_{{\bf g_\parallel}} \left( \int dz'\rho(z',{\bf g_\parallel})
\frac{2 \pi}{|{\bf g_\parallel|}} \
e^{-|{\bf g_\parallel}|\ |z-z'|} \right)
e^{-i{\bf g_\parallel}\cdot {\bf r_{\parallel}}}
\nonumber\\
\equiv & \sum_{{\bf g_\parallel}} V^{(C)}(z,{\bf g_\parallel})\
e^{-i{\bf g_\parallel}\cdot {\bf r_{\parallel}}}
\label{Coul}
\end{align}
where
\[
\rho(z,{\bf g_\parallel})  = \frac{1}{A}\ \iint d^2 {\bf r_{\parallel}}
\ \rho({\bf r})
\ e^{i{\bf g_\parallel}\cdot {\bf r_{\parallel}}} \ .
\]
\subsubsection{USPP scheme}
For Vanderbilt's ultra-soft pseudopotentials \cite{DV, LPCLV},
Eq. (\ref{KHeq}) becomes a secular equation of the form
\begin{equation}
H|\phi_i>=\epsilon_iS|\phi_i>
\end{equation}
under a generalized orthonormality condition
\begin{equation}
<\phi_i|S|\phi_j>=\delta_{ij}\ .
\end{equation}
$S$ is a Hermitian overlap operator given by
\begin{equation}
S=I+ \sum_{\sigma,{\bf R^\sigma}}\sum_{nlmn'l'm'}
q_{nlm,n'l'm'}^{\sigma}
|\beta_{n,lm}^{\sigma,{\bf R^{\sigma}}}>
<\beta_{n',l'm'}^{\sigma,{\bf R^{\sigma}}}|\ ,
\end{equation}
where $q_{nlm,n'l'm'}^{\sigma}=
\iiint d^3{\bf r} Q_{nlm,n'l'm'}^{\sigma}({\bf r})$.
Here, $\beta_{n,lm}^{\sigma}$ and $Q_{nlm,n'l'm'}^{\sigma}$ vanish outside
the core region.

The individual atomic nonlocal potential in Eq. (\ref{nlcry}) is modified as
\begin{equation}
V_{\text{nl}}^{\sigma}({\bf r}-{\bf R^{\sigma}}) =
\sum_{nlmn'l'm'} E_{nlm,n'l'm'}^{\sigma,{\bf R^{\sigma}}}
|\beta_{n,lm}^{\sigma,{\bf R^{\sigma}}}>
<\beta_{n',l'm'}^{\sigma,{\bf R^{\sigma}}}|\ ,
\end{equation}
with
\begin{equation}
E_{nlm,n'l'm'}^{\sigma,{\bf R^{\sigma}}}=
E_{nl,n'l'}^{0,\sigma}\delta_{l,l'}\delta_{m,m'}+
\iiint d^3 {\bf r}V_{eff}({\bf r})Q_{nlm,n'l'm'}^{\sigma}
({\bf r}-{\bf R^\sigma}) \ .
\label{Dval}
\end{equation}
$V_{eff}$ is defined as
\begin{equation}
V_{eff}= V_{H}+ V_{xc}+ V_{\text{LOC}}=V^{scf}+ V^{'}_{\text{LOC}}\ .
\end{equation}
Following Ref. \cite{LPCLV}, we define a box, which is large enough to
contain the core region.
The USPP $\beta_{n,lm}^{\sigma}(z-R_z^\sigma,{\bf r_\parallel})$ 
inside the box is transferred
to ${\bf G_\parallel}$ space using the fast Fourier transform (FFT), then
\begin{equation}
\beta_{n,lm}^{\sigma,{\bf R^{\sigma}}}=
\beta_{n,lm}^{\sigma}({\bf r} - {\bf R^\sigma})=
\sum_{\bf G_\parallel} \beta_{n,lm}^{\sigma}(z-R_z^\sigma,{\bf G_\parallel})
\ e^{-i\boldsymbol{\bf G_\parallel}
\cdot {\bf R_\parallel^\sigma}}
\ e^{i\boldsymbol{\bf G_\parallel}\cdot {\bf r_{\parallel}} }\ .
\end{equation}
So, the projection of $ \beta_{n,lm}^{\sigma}({\bf r} - {\bf R^{\sigma}})$
on the basis in Eq. (\ref{inpro}) reads as
\begin{multline}
<\ {\bf k_\parallel} + {\bf G_\parallel} ; i, \kappa
|\beta_{n,lm}^{\sigma,{\bf R^{\sigma}}} >  \\
= \ e^{-i\boldsymbol{({\bf G_\parallel}+{\bf k_\parallel})} \cdot
{\bf R_\parallel^\sigma} }
\sum_{\bf G_\parallel^{'}} \int dz \ B_{i;\kappa}(z)\
\beta_{n,lm}^{\sigma}(z-R_z^\sigma,{\bf G_\parallel^{'}})
\iint d^2 {\bf r_{\parallel}}\
\ e^{i\boldsymbol{({\bf G_\parallel^{'}}-{\bf G_\parallel}-{\bf k_\parallel})}
\cdot {\bf r_{\parallel}}}\ . \label{proj}
\end{multline}
Presently, we restrict ourselves to
\begin{equation}
Q_{nlm,n'l'm'}^{\sigma}({\bf r})=Q_{nl,n'l'}^{\sigma}(r)Y_{LM}
\end{equation}
with $L$ be the minimum $l_{min}$ of $|l-l'|$ and $|M=m+m'|\leq l_{min}\ .$
Similarly, both $ Q_{nlm,n'l'm'}^{\sigma}({\bf r}-{\bf R^\sigma})$ and
$V_{eff}({\bf r})$ in Eq. (\ref{Dval}) are also transferred by FFT,
\begin{equation}
Q_{nlm,n'l'm'}^{\sigma}({\bf r}-{\bf R^\sigma})=
\sum_{\bf g_\parallel} Q_{nlm,n'l'm'}^{\sigma}(z-R_z^\sigma,
{\bf R_\parallel^\sigma};
{\bf g_\parallel})
\ e^{i\boldsymbol{\bf g_\parallel}\cdot {\bf r_{\parallel}} }\ , \label{fft1}
\end{equation}

\begin{equation}
V_{eff}({\bf r})=
\sum_{\bf g_\parallel} V_{eff}(z,{\bf g_\parallel})
\ e^{i\boldsymbol{\bf g_\parallel}\cdot {\bf r_{\parallel}} } \ . \label{fft2}
\end{equation}
Then, we obtain
\begin{equation}
E_{nlm,n'l'm'}^{\sigma,{\bf R^\sigma}}=
E_{nl,n'l'}^{0,\sigma}\delta_{l,l'}\delta_{m,m'}+
A_b^{\sigma} \sum_{\bf g_\parallel} \int dz \left [ V_{eff}(z,{\bf g_\parallel})
\right ]^*
Q_{nlm,n'l'm'}^{\sigma}(z-R_z^\sigma,{\bf R_\parallel^\sigma};
{\bf g_\parallel}) \ . \label{eval}
\end{equation}
$A_b^{\sigma}$ is the surface area of the core region box.
Note that the FFT grid density for ${\bf g_\parallel}$
in the summation of Eqs. (\ref{fft1})
and (\ref{fft2}) is not necessarily the same with
that for the wavefunction \cite{LPCLV}.
$E_{nlm,n'l'm'}^{\sigma,{\bf R^\sigma}}$ in the above should
be calculated self-consistently.
From Eqs. (\ref{proj}) and (\ref{eval}), we can evaluate  
$V_{\text{NL}}|\phi_i>$ and $S|\phi_i>$.

Finally, the charge density from the wave function is augmented inside the
core region,
\begin{equation}
\rho_e({\bf r}) =\sum_i [\ |\phi_i({\bf r})|^2 +
\sum_{\sigma,{\bf R^\sigma}}\sum_{nlmn'l'm'}
Q_{nlm,n'l'm'}^{\sigma}({\bf r}-{\bf R^\sigma})
<\phi_i|\beta_{n,lm}^{\sigma,{\bf R^{\sigma}}}>
<\beta_{n',l'm'}^{\sigma,{\bf R^{\sigma}}}|\phi_i>\ ].
\end{equation}

These formula can be easily extended for the 1D case, i.e., using two sets of
B-splines to describe the non-periodic directions and 1D plane waves
for the periodic one.

\subsection{Generalized Lanczos-Krylov method for diagonalization}
In most cases, the only practical approach
to find the lowest eigenvectors of the Hamiltonian
matrix $H$ with a very large dimension
is through iterations.
In particular, it is known \cite{DS} that the use of Krylov subspaces
provides stability for the iteration process.

We start with a diagonalization in the subspace of the union
of all the input trial vectors with the Lanczos process \cite{Lan}
and the Krylov subspace $K$ generated by
repeated operations of $H$ on one of the trial vectors
${\bf v}$:
\[
K = \text{Span}\{ {\bf v}, H {\bf v}, H^2 {\bf v}, \cdots H^k {\bf v} \} \ .
\]
The orthonormality of the vectors is maintained throughout
by the standard Gram-Schmidt orthogonalization procedure. 
Thus a set of normalized roots arranged according to their eigenvalues
are obtained.

In the next step, these roots are divided into blocks of small sets of
the trial vectors. In each block,
the diagonalization is then performed in the subspace of
these vectors together with a Krylov subspace $K$
based on one of the trial vectors in the block.
Again the orthonormality of the vectors is maintained throughout
by the Gram-Schmidt scheme.
Maximum overlapping with the input trial vectors in each block is used as the
criteria for the selection of the desired roots.
This criterion guarantees the stability
of the iteration process while allowing the diagonalization to be
carried out on subsets with a small number of vectors.
Additional procedures are inserted to pick up possible additional
new roots in each block.
Thus, a set of improved normalized roots in each block are obtained.

Due to its stability,
the Jacobi method is used for the diagonalization in each step.
Modern Jacobi methods can compete in speed with
Householder-based algorithms. Also, it can be easily parallelized.
An additional advantage of the Jacobi method is that it takes full advantage
of the progressive iterative approximation property of
the input eigenvectors.
This makes it especially suitable for iteration procedures.

These two alternating grand cycles are then iterated.
It is found that only a few iterations are needed.
\section{APPLICATIONS OF PRESENT METHOD} \label{cm}
To demonstrate the capability of the present method, we apply it to study 
a few examples, including the
C(001)-(2$\times$1) surface, N $\delta$-doped graphene, and
the graphene nanoribbon.
\subsection{C(001)-(2$\times$1) surface}
We first calculated the band structure of the C(001)-(2$\times$1) surface
simulated by a ten-layer slab. The relaxed positions of the surface layer are
taken from Ref. \cite{JW}, as displayed in Fig. \ref{fig3}(a).
We did not try to determine the optimum geometry
presently, which will be implemented in the future.
A total of 40 B-splines, defined over a range of 4.5 $a_0$ ($a_0=3.52~$\AA),
are used to expand the $z$-component wavefunction.
The energy cutoff $E_c$ of the 2D plane waves is 50 Ry.
The $3 \times 6$ Monkhorst-Pack grids were taken to sample the surface
Brillouin zone. We used the GTH norm-conserving
pseudopotential \cite{GTH} for the interactions between the ions and valence
electrons. The potential
$V^{scf}$ is determined self-consistently until its
change is less than $10^{-6}$ Ry. For comparison, we also performed the 
calculation by using the VASP code with the projector-augmented-wave method 
(PAW) \cite{KJ,KF}.

In practical calculations, the significant charge-density
oscillation, the so-called charge sloshing, was observed. To remedy the very
slow convergence by the simple linear mixing scheme (Fig. \ref{fig4}), we use
the Kerker mixing \cite{Kerker} to prevent the charge sloshing and the
residual minimization method in the direct inversion of iterative subspace
(RMM-DIIS) Pulay scheme \cite{CP,EPW}
to accelerate the convergence.
The new charge is mixed by
\begin{equation}
\rho_{new}(z, {\bf g_\parallel})= \rho_{in}(z, {\bf g_\parallel})+
W\frac{{\bf g_\parallel}^2}{{\bf g_\parallel}^2+{\bf g_{\parallel,max}}^2}
(\rho_{out}(z, {\bf g_\parallel})-\rho_{in}(z, {\bf g_\parallel})) \nonumber
\end{equation}
The weight factor $W$ and the cutoff wavevector ${\bf g_{\parallel,max}}$ are
two adjustable
parameters. The updated input charge density, as a linear
combination of charge density of all 4 previous steps, is determined such that
it minimizes the corresponding residual vector.

Figures \ref{fig3}(b) and (c) display the results.
The bands for wave vectors along $X'-M-X$ should be twofold degenerate due to
the symmetry of the slab system \cite{note1}.
In the present calculation, the corresponding splitting
is $<10^{-5}$ eV for all ${\bf k}$ along $X'-M-X$. 
As can be seen in Fig. \ref{fig3}(b) and (c),
our results are almost identical to those by the VASP code. The filled $\pi$
band and the empty $\pi^*$ band, separated in energy by $\sim2$ eV, are clearly 
seen in the bulk band gap, and are also in nice agreement with those in 
Ref. \cite{KP}.  

\subsection{n-type $\delta$-doped graphene}
Before proceeding to the second example, we should keep in mind that, although
the utility of first-principles norm-conserving pseudopotentials has paved
the way to accurate calculations of solid-state properties,
the norm-conserving constraint is the main factor
responsible for the hardness of highly localized
valence orbitals of $1s, 2p$, or $3d$, with no core state of the same
angular momentum. Therefore, the all-electron wave function is nodeless and
quite compressed compared to the other valence states, thus requiring a large
number of plane waves to be represented accurately.
For the next two graphene-related examples,
we will use the Vanderbilt's USPP scheme in which such a constraint is relaxed.
In this scheme, the pseudo-wave functions are
allowed to be as soft as possible within a given region, yielding a dramatic
reduction of the cutoff energy. For instance, we found that
the change in the band structure of the graphene by the NCPP will be
unnoticeable with $E_c=70$ Ry but, by the USPP with   
$E_c=20$ Ry only. This 
reduction of the cutoff energy in USPP is particularly important for large
systems to avoid the diagonalization bottleneck of the 
extremely large matrix $H$.  

First of all, we examined the quality of the C and N USPPs that were
generated from
the Vanderbilt's code \cite{vancode}, by performing the band structure
calculation for the graphene and the assumed N-doped graphene with
equal number of C and N atoms. Figure \ref{fig5} displays
the band structures, along with the VASP-PAW results. Clearly, both
are nearly identical for these two cases. Therefore, we believe that the
quality of C and N USPPs used here is sufficiently good for our band structure
calculations.

The calculations were carried out using a supercell of
$8 \times \sqrt{3}$, i.e., 16 carbon dimmer lines between adjacent N chains,
where the N chain is the $\delta$ dopant.
All C and N atoms were kept at the ideal positions, with the
lattice constant $a_0$ set to be 2.641 \AA. A mixed basis set with
13 B-splines distributed over a range of 3.25 $a_0$ and
the plane wave cutoff of 20 Ry are used.
The $2 \times 4$ Monkhorst-Pack grids were taken to sample the surface
Brillouin zone.

Overall, we found an excellent agreement between the present band structures
and the VASP-PAW results. Figure \ref{fig6} shows only the results near
the Fermi level along $\Gamma-X$.
We have also tested various N mole fractions $x$ of 1/2, 1/4, and 1/8, 
with $x$ defined as the ratio of the number of N to the total number of 
atoms per unit cell.
Besides all agreeable with the VASP-PAW results, we found that
the trend of the band gap is also in accord with previous
work \cite{WFWCZ}, e.g., the
band gap opens only for large $x$ (Fig. \ref{fig5}(c)).
Note that the USPP and PAW methods have proven to be closed
related \cite{KJ}. Therefore, we are confident that the present program has
produced reliable results.

It is worth noting that the ratio of the number of the basis $N_d$ used in  
VASP and the present approach is about 1.4 $-$ 1.7 for $E_c = 20 - 30$ Ry, with 
a typical thickness of 10 \AA~ for the vacuum layer required in VASP. 
Like in the conjugate gradient technique \cite{KF,TPA,PTA}, 
the computational effort by the present Lanczos-Krylov method 
scales as $M^2N_d$ for the orthogonalization and $MN_d$log$N_d$ for the FFT, 
where $M$ is the number of the bands considered.  
Therefore, as compared to the conventional supercell modeling, the computational
time needed for the eigenvalues search due to
the planar mixed basis will reduce with a factor of 2 - 3 if the same 
algorithm were used.  This reduction will be more significant for the cases,
in which the tail of the Coulomb potential extends to a very long distance, 
whereas the wave function still decays exponentially in the vacuum.  
\subsection{graphene nanoribbon}
Finally, we will apply the present method to
armchair graphene nanoribbon. In the spirit of our method,
we should use two sets of
B-splines describing the two non-periodic dimensions and only 1D plane waves
for the periodic $y$ direction in such a system. Here, we just adopt the
planar mixed basis set to study the nanoribbon by using the surface supercell
modeling for the $x-y$ plane.

Armchair graphene nanoribbons are characterized
by the number of dimmer lines $N_a$ (Fig. \ref{fig7}(a)) across
the ribbons. The width of the ribbon chosen is chosen to include $N_a=19$
carbon dimmer lines, which is $3p+1$ type ($p$ is a integer).
We performed the calculation using a $16 \times \sqrt{3}$ surface supercell
with the vacuum 'band' of $\sim$ 15 \AA~ along the $y$ direction. The optimized
atomic positions in the eight outer dimmer lines of both sides obtained
by the VASP code were used for the calculation.
Since graphene nanoribbons are stripes of graphene, edge
atoms are not saturated \cite{KLY}. For simplicity, we let the edge atoms not 
saturated by hydrogen atoms in the present calculation. 
All other computational conditions
are similar to the case of n-type $\delta$-doped graphene described above.

The band structures of the present USPP result and the VASP-PAW counterpart 
near the
Fermi level are displayed in Fig. \ref{fig7}(b) and (c), respectively. Again,
there is a very nice agreement between these two approaches. It has been
reported \cite{SCL} that all armchair graphene nanoribbons with the three
typical families, i.e., $N_a=3p,\ 3p+1$, and $3p+2$, are semiconductors with a
direct energy gap. The bandgap obtained by our approach is 0.57 eV, and 0.61 eV by
the VASP code. These values compare favorably with that
in Ref. \cite{SCL}.

To sum up, we are convinced that the present program has been implemented
successfully for low-dimensional systems and the results obtained are
very reliable. Moreover, as compared to traditional supercell 
modeling, the use of the present mixed basis will reduce the number of
the basis functions and speed up the calculations of electronic states.   
\section{CONCLUSIONS} \label{con}
In conclusion, we have developed an efficient and accurate method to
investigate the electronic structures of low dimensional systems
by a mixed basis set with plane waves for the periodic direction and B-spline polynomials for
the non-periodic direction. Contrary to the existing algorithms based
upon the conventional supercell model with
alternating slab and vacuum regions, it is a real space approach along
the $z$-axis and therefore
gives the surface band structures in absolute-energy scale. Furthermore,
B-splines are independent of atomic positions when the atomic
structure is optimized; thus, the geometry optimization can 
be easily implemented.
We have also implemented a generalized Lanczos-Krylov iterative method for the
diagonalization of the Kohn-Sham Hamiltonian. It offers a promising way to find the lower
eigenvalues of a real matrix with a very large dimension through iterations.
Particularly, this algorithm is very stable for low dimensional systems, of
which the wave functions vary abruptly.

We have calculated the electronic structures of
the C(001)-(2$\times$1) surface with
GTH norm-conserving pseudopotentials \cite{GTH}
and the N $\delta$-doped graphene, and graphene nanoribbon with
Vanderbilt's ultra-soft pseudopotentials \cite{DV, LPCLV}.
It is found that the band structures are all in good agreement with 
those by the popular existing codes,
but with a reduced number of basis functions.
\begin{acknowledgments}
This work was supported by the National Science Council under 
grant numbers NSC 100-2112-M-017-002-MY3 and NSC 01-2112-M-001-024-MY3 and 
by National Center for Theoretical Sciences of Taiwan.
\end{acknowledgments}
\newpage
\begin{center}
\large {\bf FIGURE CAPTIONS}    \normalsize
\end{center}
Fig. 1: Schematic plot of a real slab, investigated by (a)
        the conventional supercell model and by (b) the real space approach
        with a localized basis set.\\ \\
Fig. 2: B-splines defined on (a) an equal-distance (b) an exponential-type
        knot sequence (noted as *) of order $\kappa=5$. \\ \\
Fig. 3: (Color online) (a) Atomic structure of the C(001)-(2$\times$1)
        surface. Atoms located in deeper layer are denoted by smaller circles.
        (b) and (c) are the corresponding band structures obtained by
        the present work with NCPP and VASP with PAW. \\ \\
Fig. 4: (Color online) Convergence rate for the calculation of the
        C(001)-(2$\times$1) surface
        by the Pulay-Kerker mixing and simple linear mixing scheme. \\ \\
Fig. 5: Band structures of graphene obtained by 
        (a) the present work with USPP and (b)
        VASP with PAW. (c) and (d) are the band structures of 
        N $\delta$-doped graphene obtained by the
        present work with USPP and VASP with PAW, respectively. \\ \\
Fig. 6: (Color online) (a) Atomic structure of the N $\delta$-doped graphene.
	Red circles denote N atoms.
        (b) and (c) are the corresponding band structures by
        the present work with USPP and VASP with PAW. \\ \\
Fig. 7: (Color online) (a) Atomic structure of the armchair graphene
        nanoribbon. (b) and (c) are the corresponding band structures by
        the present work with USPP and VASP with PAW. \\ \\

\newpage

\newpage
\newpage
\begin{figure}[h]
        \caption{ } \label{fig2}
\includegraphics{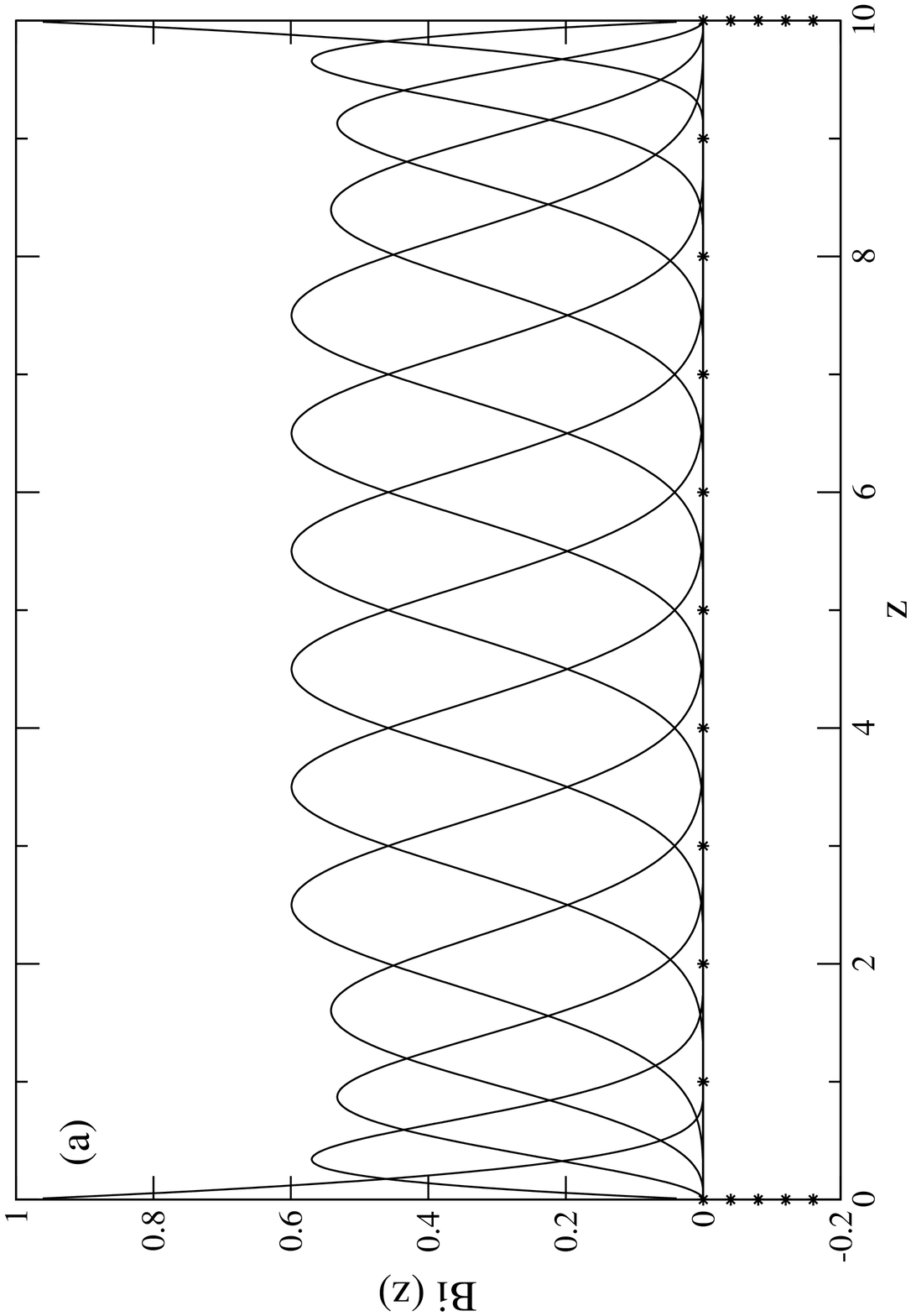}
\includegraphics{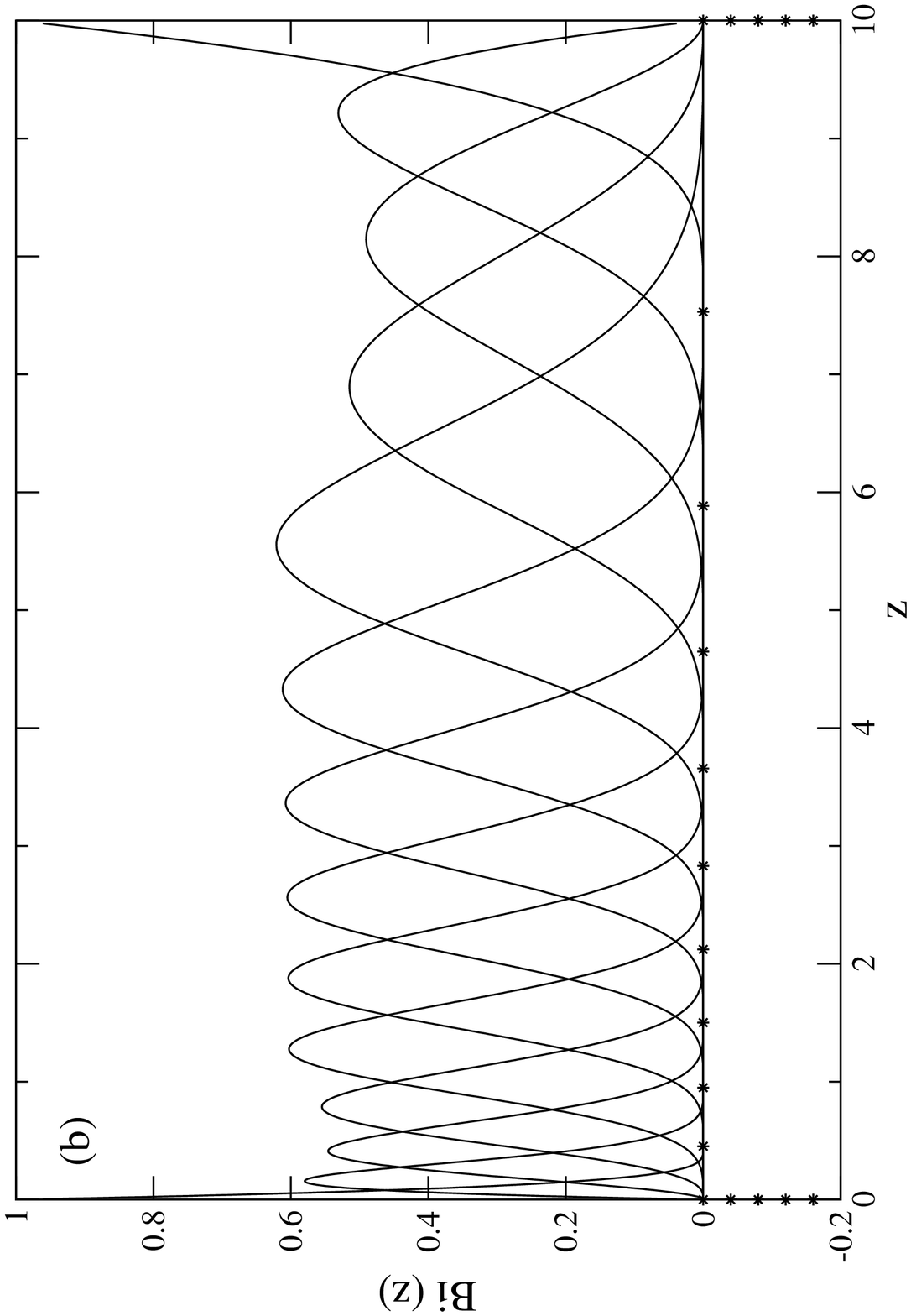}
\end{figure}
\newpage
\begin{figure}[h]
        \caption{ } \label{fig3}
\includegraphics{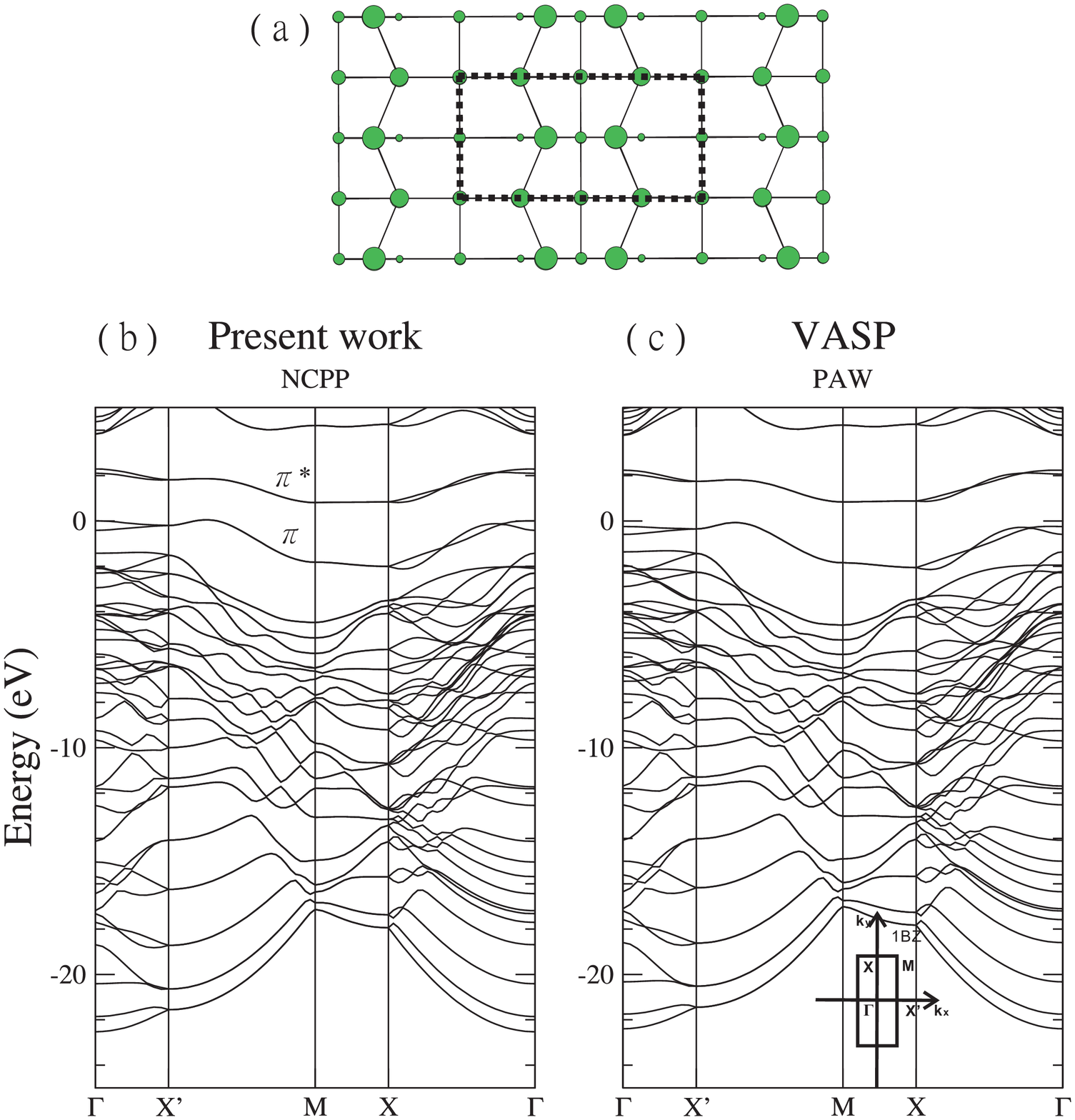}
\end{figure}
\newpage
\begin{figure}[h]
        \caption{ } \label{fig4}
\includegraphics{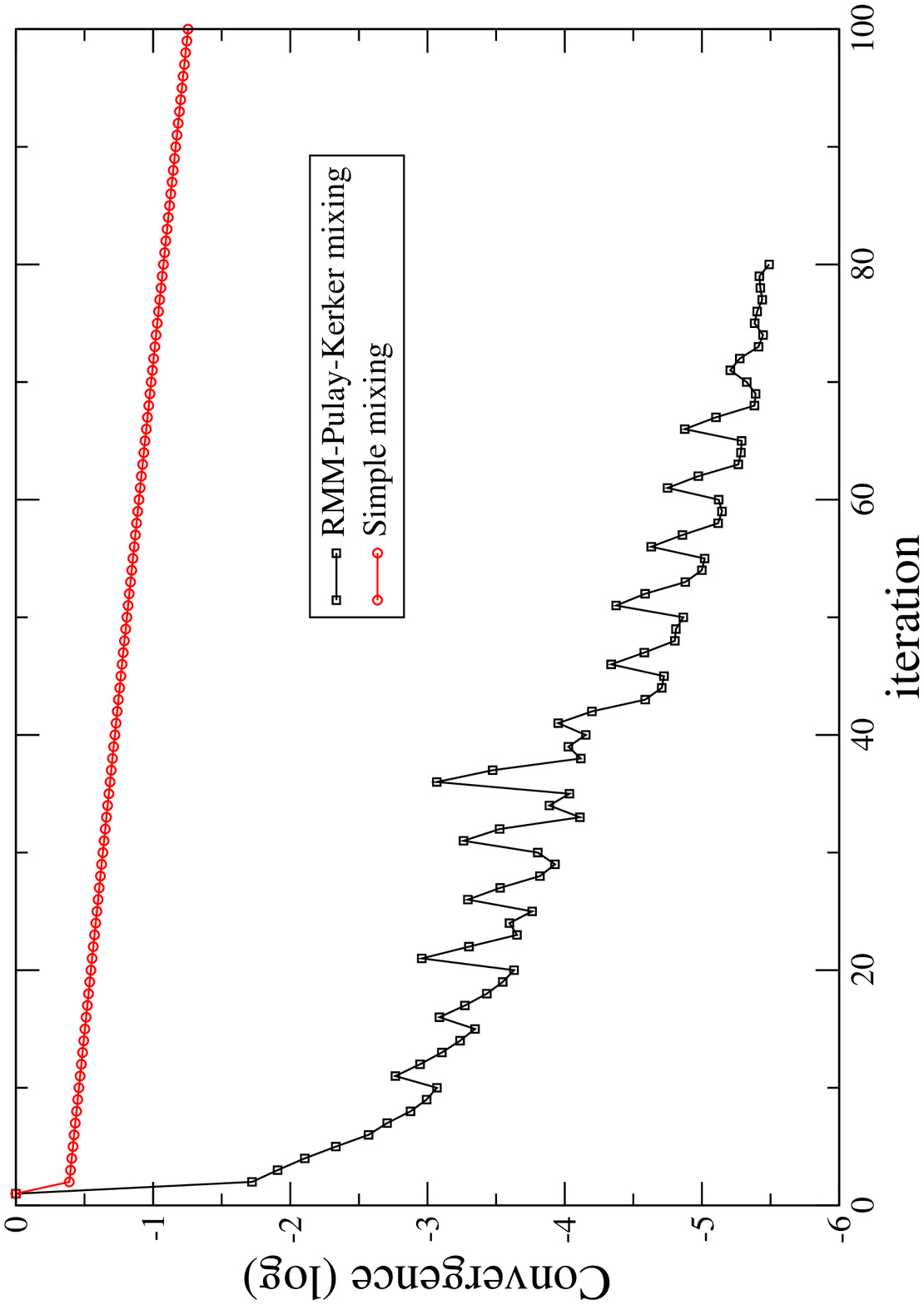}
\end{figure}
\newpage
\begin{figure}[h]
        \caption{ } \label{fig5}
\includegraphics{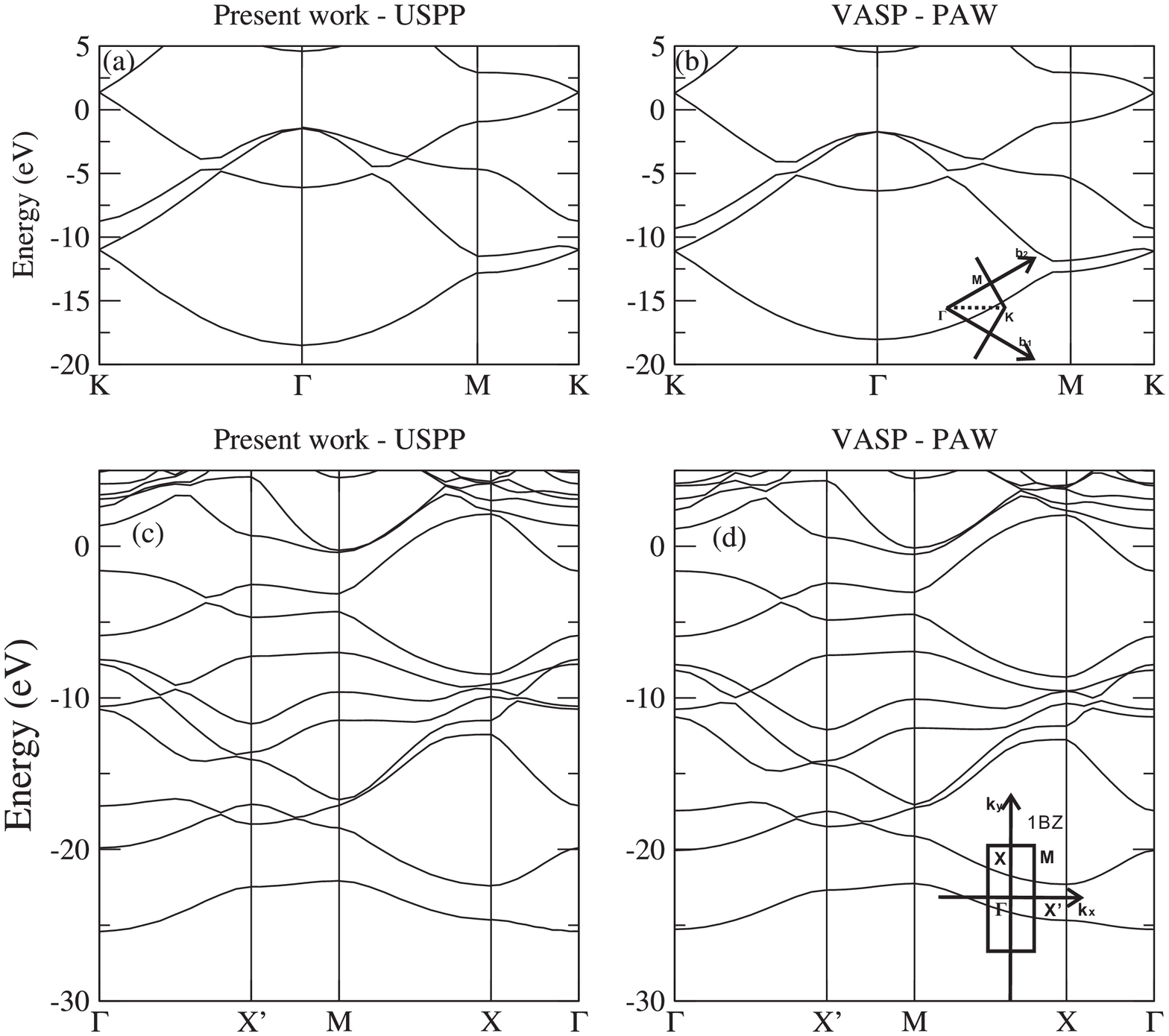}
\end{figure}
\newpage
\begin{figure}[h]
        \caption{ } \label{fig6}
\includegraphics{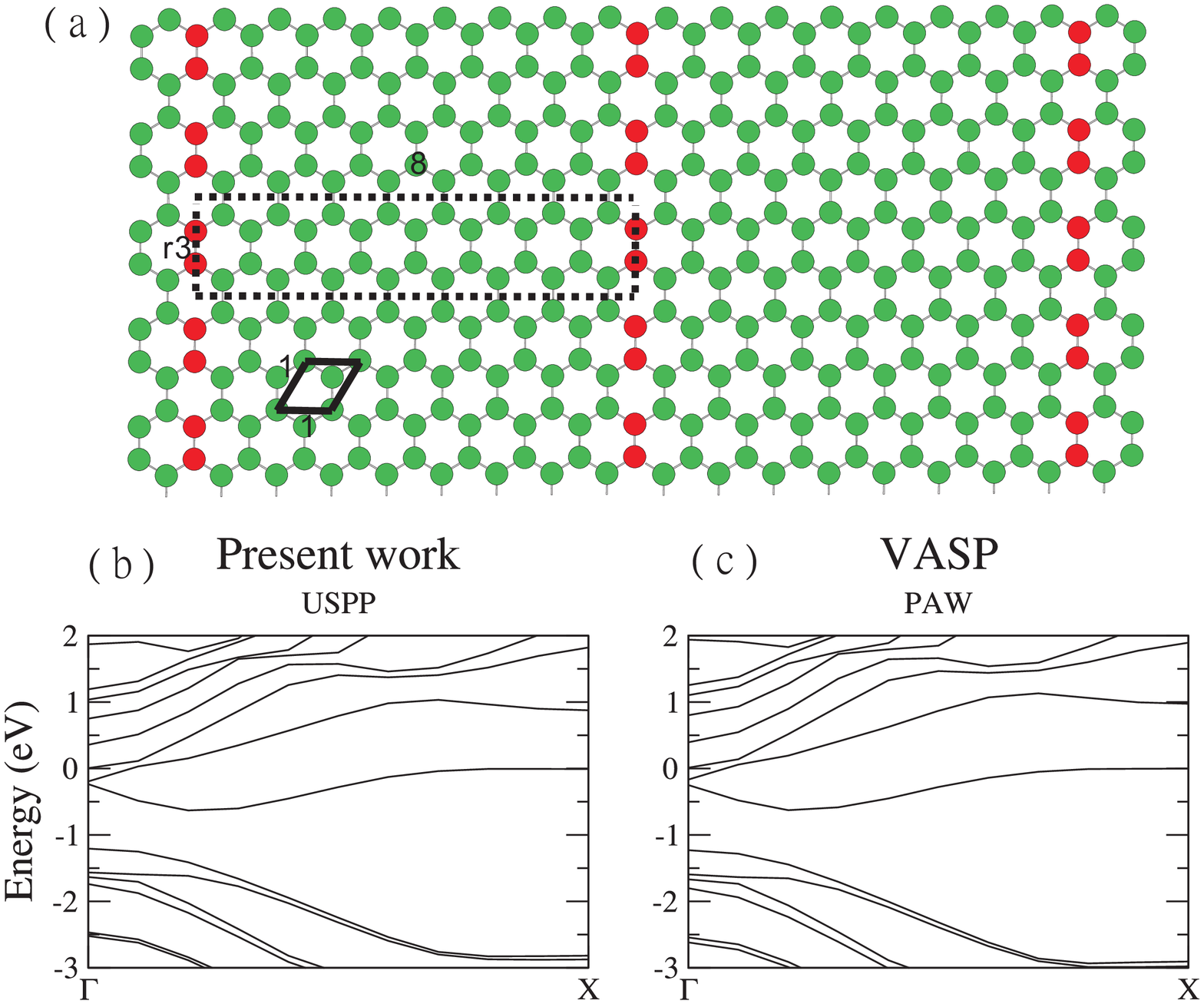}
\end{figure}
\newpage
\begin{figure}[h]
        \caption{ } \label{fig7}
\includegraphics{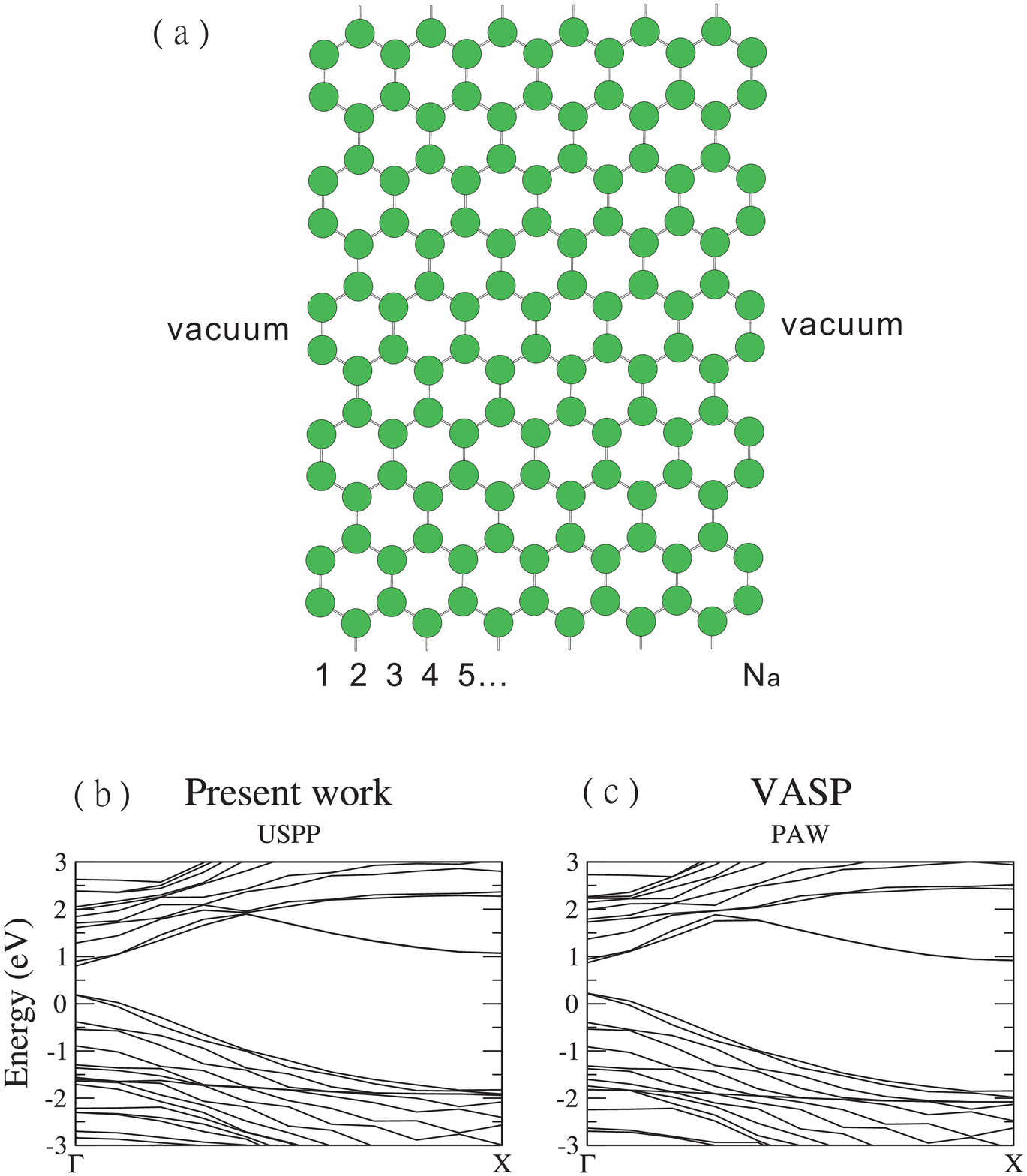}
\end{figure}

\begin{thebibliography}{99}
\bibitem{note1}
Here, we use the symmetric dimer model for C(001)-(2$\times$1) reconstrued
surface for testing the present approach.
\end{thebibliography}
\end{document}